\documentclass[12pt]{article}

\usepackage{latexsym}

\begin{document}

\title{Classical Liouville Theory and the Microscopic Interpretation of Black Hole Entropy }
\author{A. Giacomini\thanks{e-mail: giacomin@science.unitn.it}\\ Dipartimento di Fisica Universit\`a di Trento\\
 and Istituto Nazionale di Fisica Nucleare}
\date{}
\maketitle

\begin{abstract}
In this paper we will compute the entropy of a Schwarzschild black hole by finding a classical central charge 
of the Virasoro algebra of a liouville theory and using the Cardy formula. This is done by performing a dimensional reduction 
of the Einstein-Hilbert action with the Ansatz of spherical symmetry and writing the metric in conformally flat form. We obtain
two coupled field equations and using the near horizon approximation the field equation for the conformal factor decouples becoming
a Liouville equation. The generators of conformal transformations of a liouville theory form a Virasoro algebra with a classical central charge.
It is then possible to compute the black hole entropy via Cardy formula and so to count the microstates responsible for this entropy.
This computation is independent from a specific quantum theory of gravity model.  
\end{abstract}

\section{Introduction}
It is a well known fact that for black holes there are 4 theorems of differential geometry called the ``4 laws of black hole mechanics''
\cite{Giacomini:hawking&bardeen&carter}.  
\begin{equation}
\kappa = const \; \;  \mathrm{ on \, H} \label{Giacomini:law1} 
\end{equation}
\begin{equation}
\delta M = \frac{\kappa}{8\pi} \delta A + \Omega \delta J +\Phi \delta e \label{Giacomini:law2}
\end{equation}
\begin{equation}
\delta A = \geq 0 \; \;  \mathrm{in \,  every \,  phys. \, process} \label{Giacomini:law3}
\end{equation}
\begin{equation}
\kappa =0 \; \;  \mathrm {BH \,  not \,  realizable} \label{Giacomini:law4}
\end{equation}
One can notice the strong analogy to the laws of thermodynamics if one associates to the mass $M$  the energy to the surface gravity $\kappa$
the temperature and to to the horizon area $A$ the entropy. But at classical level this is only a formal analogy because a black hole being ``black''has 
$T=0$. The situation changes dramatically when one considers quantum effects near the horizon. Hawking
\cite{Giacomini:hawking} showed in his famous article that due to 
quantum effects a black hole emits a thermal radiation proportional to the surface gravity
\begin{equation}
T=\frac{\kappa}{2\pi} \; . \label{Giacomini:rad1}
\end{equation}
So the 4 laws of black hole mechanics are not only an analogy to the laws of thermodynamics but are really the laws of thermodynamics applied
to black holes. So one can conclude that a black hole possesses an entropy given by the Bekenstein-Hawking formula \cite{Giacomini:bekenstein}
\begin{equation}
S=\frac{A}{4} \; .   \label{Giacomini:entro1}
\end{equation}
Now in thermodynamics the entropy has a statistical interpretation in terms of microstates given by the boltzmann formula
\[
S=k_B \log N \; .
\]
So  there arises the question wich are the microstates of the black hole responsible for the entropy.
One possibility would be, if we consider the black hole as a collapsed star, to take the degrees of freedom of matter as responsible
for the entropy. But if we compare the entropy of a non collapsed star of 1 solar mass that is about
$S=10^{58} k_B$ with the entropy of a black hole of 1 solar mass that is abot $S=10^{77} k_B$ we see that there are 19 magnitudes missing.
One can conclude that the relevant role for the black hole entropy is played by the gravitational degrees of freedom.
The problem is now that no complete quantum theory of gravty exists at the moment. There exist of course some specific models for
quantum gravity like superstrings or loop quantum gravity, wich can succesfully compute the entropy of some classes of black holes,
but it remains the fact that the black hole entropy was found by means of semiclassical calculations and already at classical level
one can see the thermodynamic properties of black holes by means of the 4 laws of black hole mechanics.
We will therefore show in this paper ,which is based on an article written by A. Giacomini \& N. Pinamonti, \cite{Giacomini:giacomini&pinamonti} that it is possible
to count the microstates of a black hole in a way that is completely independent of a specific model of quantum theory of gravity.
This is done by finding a classical symmetry principle, in our case the 2-D conformal symmetry.\\
In order to understand the role of 2-D conformal symmetry in counting the black hole microstates let us recall some 
standard features of 2-D conformal symmetry.
Introducing complex coordinates the flat metric can be written in the form 
\[
ds^2= dzd\overline{z}
\]
and we see that every analytic function 
\[
z\rightarrow f(z) \; \; ; \; \;\overline{z} \rightarrow \overline{f}(\overline{z})
\]
defines a conformal transformation
\[
ds^2 = \frac{\partial f}{\partial z} \frac{\partial \overline{f}}{\partial \overline{z}}=\rho (z\overline{z})dzd\overline{z} \; .
\]
The infinitesimal generators of such transformations 
\[
G_n=z^{n+1} \partial _z
\]
close a lie algebra
\begin{equation}
[G_n , G_m] =(n-m)G_{n+m} \label{Giacomini:lie1} \; .
\end{equation}
In 2D CFT the stress energy tensor has only two components i.e. one analytic and the other antianalytic
\begin{equation}
T= T_{zz} \; \; ;  \; \;\overline{T}= \overline{T}_{\overline{zz}} 
\end{equation}
\begin{equation}
\partial _z \overline{T} =0 \; \; ; \; \; \partial _{\overline{z}} T =0
\end{equation}
We can therefore expand for example $T$ in Laurent series 
\[
T(z)=\sum _{-\infty} ^{\infty} \frac{L_n}{z^{n+2}} \; .
\]
In the quantum case the $L_n$ become operators and relatively to the commutator form a Virasoro algebra
\begin{equation}
[L_n , L_m]= (n-m)L_{n+m} + \frac{c}{12}(m^3 -m)\delta _{m, -n} \; .  \label{Giacomini:virasoro1}
\end{equation}
This is a central extension of the algebra (\ref{Giacomini:lie1}). The central extension in the commutator algebra arises because of the
normal ordering of the creation operators.
It is a standard result for quantum CFT in 2-D that the asymptotic density of states for given $L_0$ is completely determined by
the Virasoro algebra by means of the  Cardy formula \cite{Giacomini:cardy}
\begin{equation}
\rho(L_0 ) =\exp \left( 2\pi \sqrt{\frac{cL_0}{6}} \right) \; .  \label{Giacomini:cardy1} 
\end{equation}
The entropy can therefore be calculated by using the logarithm of the Cardy formula
\[
S= \ln \rho \; .
\]
A central extension of  the conformal algebra (\ref{Giacomini:lie1}) can already arise at classical level in the Poisson algebra of the charges,
as for example in the Liouville theory \cite{Giacomini:jackiw}. \\
How can we now use the powerful tools of 2-D CFT for the counting of microstates of a black hole?
We notice that all the relevant geometry of a black hole is encoded in the $r-t$ plane which is a 2-D manifold .
Therefore if we find a  classical conformal symmetry for the geometry of the $r-t$ plane which admits a virasoro algebra with
a classic central charge we would be able to count the microstates of the black hole via Cardy formula without the use of 
any quantum theory of gravity. One interesting approach in this direction was made by Carlip \cite{Giacomini:carlip} computing the Poisson
brackets of the canonical generators which preserve certain fall off conditions of the metric near the horizon.
This idea is based on an article of J.D. Brown and M. Henneaux, where the central charge of the Poisson algebra  of generators 
preserving the asymptotic structure of $AdS_3 $ is found \cite{Giacomini:brown&henneaux} .
This approach seems to have some technical difficulties especially for the Schwarzschild black hole \cite{Giacomini:park} , \cite{Giacomini:soloviev}.
We will use in this paper a different approach: using the Ansatz of spherical symmetry (we are interested in the Schwarzschild
 black hole) we will perform a dimensional reduction
and obtain an effective 2-D theory. Using then a near horizon approximation  we will eventually find that all the dynamics is 
described by a Liouville theory with a computable classical central charge. We can then use this classical Virasoro algebra to
compute the Schwarzschild  black hole entropy via Cardy formula.

\section{Dimensional reduction}
Let us start with the usual 4-dimensional Einstein-Hilbert action

\begin{equation}
I= \frac{1}{16\pi}\int \sqrt{-g} R d^4 x \; .     \label{Giacomini:einstein1}
\end{equation}
We make the Ansatz of spherical symmery

\begin{equation}
ds^2= g^{(2)}_{ij}dx^i dx^j + \Phi^2 (x_1 ,  x_2) \left( d\theta^2  + \sin ^2 \theta d\phi ^2  \right) \; ,        \label{Giacomini:spheric1}
\end{equation}
where $g^{(2)}$ is the metric of  a 2-D manifold and $\Phi$ is a radial coordinate.
We write now the action (\ref{Giacomini:einstein1}) in function of the 2-dimensional curvature scalar $R^{(2)}$ and $\Phi$ and integrate over the angular 
variables. We obtain  an effective action of a  2-dimensional gravity coupled to a dilaton field 

\begin{equation}
I=\frac{1}{4} \int  d^2 x \sqrt{-g^{(2)}}\left( 2(\nabla \Phi)^2 +\Phi ^2 R^{(2)} +2  \right)  \; .       \label{Giacomini:reduced1}
\end{equation}
One possibility would be to study the dynamics of the dilaton field and see if near horizon it becomes a conformal field.
This has already been done by Solodukhin \cite{Giacomini:solodukhin}.
In this paper we argue that the relevant degree of freedom is the metric of the $r-t$ plane and so
in order to put this action in a more useful form we redefine

\begin{equation}
\Phi^2 = \eta \; \; \; ;  \; \; g_{ab} ^{(2)} = \frac{1}{\sqrt{\eta}} \tilde{g}_{ab}     \label{Giacomini:redefine1}
\end{equation}
obtaining the action

\begin{equation}
I=\frac{1}{2} \int \sqrt{-\tilde{g}} \left[ \frac{\eta}{2} R[\tilde{g}] + V(\eta )   \right] \; ,  \label{Giacomini:reduced2}
\end{equation}

where the dilatonic potential  $V(\eta )$ in our case is $V(\eta) = \frac{1}{\sqrt{\eta}}$ .
Up to now we have a theory with 2 degrees of freedom: The field $\eta$ which is related to the radius of the 2-sphere  while 
$\tilde{g} $ contains the geometry of the $r-t$ plane. 
It is well known that in 2 dimensions the metric can always be written in conformally flat form
 
\begin{equation}
\tilde{g} _{ab} = e^{-2\rho} \gamma _{ab}    \; ,    \label{Giacomini:conformflat1}
\end{equation}
where  $\gamma$ is the 2-dimensional Minkowski metric. Therefore all the geometry of the $r-t$ plane is described by the field
$\rho$. The action (\ref{Giacomini:reduced2}) using (\ref{Giacomini:conformflat1}) becomes

\begin{equation}
I= \frac{1}{2} \int d^2 x \left( -\partial _a \eta \partial ^a \rho + V(\eta ) e^{-2\rho}   \right)  \; .     \label{Giacomini:reduced3}
\end{equation}
We obtain so a theory of two fields propagating in flat spacetime. From the last action we obtain two coupled equations of motion

\begin{equation}
\Box \rho + \partial _{\eta} V(\eta ) e^{-2\rho} =0 \; \; ;  \; \;   \Box \eta - 2V(\eta ) e^{-2\rho}=0  \; .      \label{Giacomini:motion1}
\end{equation}
As we have gauge fixed the metric to the form (\ref{Giacomini:conformflat1}) the fields must also satisfy the constraints

\begin{equation}
\frac{\delta I }{ \delta g^{ab}} = T_{ab}=0     \; .             \label{Giacomini:constraint1}
\end{equation}

Introducing now Light coordinates $ x^{\pm} = x^1 \pm x^2 $  the equations of motion (\ref{Giacomini:motion1}) take the form 
 
\begin{equation}
\partial _{+} \partial _{-} \rho - \frac{\partial_{\eta} V(\eta )}{4} e^{-2\rho}=0 \; \; ;  \; \; \partial _+ \partial _- \eta
+ \frac{V(\eta )}{2} e^{-2\rho} = 0                                  \label{Giacomini:motion2}
\end{equation}
and the constraints (\ref{Giacomini:constraint1} ) can now be written as 

\begin{equation}
T_{\pm \pm } =\partial _{\pm} \partial _{\pm} \eta + 2 \partial _{\pm} \rho \partial _{\pm} \eta =0 \; .    \label{Giacomini:constraint2}
\end{equation}
Notice that the last equation of (\ref{Giacomini:motion2}) plus the constraints imply the first equation of (\ref{Giacomini:motion2}).

\section{Near-horizon approximation}
 
Up to now we have only made  an Ansatz of spherical symmetry. If we impose the existence of a black hole the $r-t$ plane metric
becomes

\begin{equation}
ds^2 = -N^2 (r)dt^2 + \frac{1}{N^2 (r)} dr^2 \; .        \label{giacomini:bhmetric1}
\end{equation}
Now as near horizon approximation the lapse function $N$ can be approximated by

\begin{equation}
N^2 (r) = 2\kappa (r-r_0 )   \; .    \label {Giacomini:lapse1}
\end{equation}
Introducing the coordinate $r=r_0 +\kappa y^2 /2 $ the near horizon metric takes the Rindler  form

\begin{equation}
ds^2 = -\kappa ^2 y^2 dt^2 + dy^2          \label{Giacomini:rindler1}
\end{equation}
Now introducing $x^{\pm} = \kappa t \pm \log y $ we eventually obtain

\begin{equation}
ds^2 = -dx^+ dx^- \exp \left(  x^+ - x^- \right) \; .          \label{Giacomini:rindler2}
\end{equation}
Comparing this expression with the equation (\ref{Giacomini:conformflat1}) we can find the form of the conformal factor $\rho$ near the horizon

\begin{equation}
-2\rho = x^+ - x^-     \; .      \label{Giacomini:factor1}
\end{equation}
Notice that it easy to integrate the constraint (\ref{Giacomini:constraint2}) 

\begin{equation}
\partial_{\pm} \eta = \exp \left( -2\rho +C_{\mp}(x_{\mp}  \right) \; ,      \label{Giacomini:integrated1}
\end{equation}
where $ C_{\mp} $ is an arbitrary function of $ x^{\mp} $
In the near horizon limit i.e. taking $x^{\pm} \rightarrow \mp \infty $ and because of  (\ref{Giacomini:factor1})
this means $\rho\rightarrow \infty $ we obtain

\begin{equation}
\partial_{\pm} \eta = 0 \Rightarrow \eta =const  \; .                        \label{Giacomini:constant}
\end{equation}  
This means that in the near horizon limit the dilaton field can be considered as fixed at it's value on the horizon $\eta_0$.
In that case it is immediate to check that the constraints (\ref{Giacomini:constraint2}) and the second equation of motion in 
(\ref{Giacomini:motion2}) are identically satisfied. As equation of motion survives therefore only

\begin{equation}
\partial _+ \partial_- \rho  + \frac{1}{8\eta _0 ^{3/2}} e^{-2\rho}  =0   \; .           \label{Giacomini:liouville1}
\end{equation}
This is the classical Liouville equation. Therefore we can conclude that in the near horizon limit all the dynamics of the black
hole is described by the field $\rho$, which describes the geometry of the $r-t$ plane. The dynamics of this field is governed by the Liouville 
equation.
It is a well known fact that the Liouville theory possesses a clasical central charge.

\section{Virasoro algebra}

In order to find a Virasoro algebra we must introduce a set of generators. To do this let us first notice that the most general action
that gives the Liouville equation is

\begin{equation}
I = C\int \sqrt{\hat{g}} \left( \frac{1}{2} \partial_{\mu} \rho \partial^{\mu} \rho + \frac{\mu}{\beta ^2}e^{-\beta \rho} - 
\frac{2}{\beta} \rho R[\hat{g}]  \right)    \; .                  \label{Giacomini:liouvaction1}
\end{equation}
In our case $\beta =2$ and $\mu = 1/ \eta ^{3/2}$. It is important to stress that we haven't derived the action (\ref{Giacomini:liouvaction1})
from the gravitational one (\ref{Giacomini:reduced3}) but this is the most general action that gives the Liouville equation
(\ref{Giacomini:liouville1}). The constant $C$ in the action is arbitrary for the moment and will be fixed later.
The stress energy tensor derived from the last action is in lightcone coordinates 

\begin{equation}
T_{\pm \pm} = C \left( \partial _{\pm} \rho \partial _{\pm} \rho + \frac{2}{\beta}\partial _{\pm} \partial_{\pm} \rho  \right) \; . 
                                                                      \label{Giacomini:liouvstress1}
\end{equation}
The term with the second derivatives arises from the variation of the scalar curvature term in the action (\ref{Giacomini:liouvaction1})
and remains also after choosing a flat fixed background , where $R$ is then of course zero.
Now the constant $C$ in the action (\ref{Giacomini:liouvaction1}) is set in such a way that the energy of the system equals
the mass of the black hole $M_B$.

\begin{equation}
\int_{-l/ \sqrt{2}} ^{l/ \sqrt{2}} T_{11} dx^2 =M_B \Rightarrow C= \frac{\kappa A }{2\pi l} \; .         \label{Giacomini:costant1}
\end{equation}
The parameter $l$ is a cutoff parameter that eventually must tend to infinity.
Now we can define the generators

\begin{equation}
L_n ^{\pm} = \int _{-l/ 2\kappa} ^{l/ 2\kappa} \kappa dx^{\pm} \xi_n ^{\pm} T_{\pm \pm}  \; ,         \label{Giacomini:viragen1}
\end{equation}
where the factor $2\kappa$ in the integration is used to match the euclidean periodicity. 
The $\xi_n$ are smearing fields defined as
 
\begin{equation}
\xi_n ^{\pm} = \frac{l}{2\kappa \pi}\exp \left( -i \frac{2\pi \kappa}{l} n x^{\pm}  \right)   \; .           \label{Giacomini:xi1}
\end{equation}
We can now compute easily the Poisson brackets of the generators
\begin{equation}
\left\{ L_n ^{\pm} , L_m ^{\pm}  \right\} _{PB} =i(n-m)L_{m+n} ^{\pm} + i \frac{c}{12}n^3 \delta _{m+n ,0} \; ,       \label{Giacomini:Virasoro}
\end{equation}

with 

\begin{equation} 
L_0 ^+ = \frac{A l}{16\pi ^2 }        \label{Giacomini:zeromode1} 
\end{equation}
\begin{equation}
c= \frac{3A}{2l}       \; .     \label{giacomini:charge1}
\end{equation}
Now for every fixed value of the cutoff $l$ we are able to compute the entropy with the logarithm of the  Cardy formula
 
\begin{equation}
S= 2\pi \sqrt{\frac{c^+ L_0^+}{6} } = \frac{A}{4}  \; ,        \label{Giacomini1:bekenstein1}
\end{equation}
which is exactly the Bekenstein-Hawking entropy.  If we let tend the cutoff to infinity the $L_0$ generator diverges and the central charge tends to 
zero. But we notice in the Cardy formula we have the product $L_0 c$ and therefore the result of the computation does not depend on
$l$. We remember that a zero central charge was also found in \cite{Giacomini:koga} , \cite{Giacomini:hotta&sasaki}. 
Therefore evntually we have ``counted'' the microstates of a black hole without using detail of a specific model of quantum theory of gravity,
but using only the classical near horizon structure of the black hole.

\subsection*{Acknowledgements}

The author wants to thank N. Pinamonti for  his cooperation in the research on the subject.
The author is also grateful to L. Vanzo , V. Moretti , S. Zerbini and M. M. Caldarelli for useful discussions on the subject.

\end{document}